\documentclass[letter]{jpconf}
\usepackage{graphicx}
\usepackage{url}
\begin{document}
\title{Hadro-production measurements for T2K by NA61/SHINE at the CERN SPS}

\author{Claudia Strabel for the NA61/SHINE Collaboration \url{http://na61.web.cern.ch/}}

\address{Institute for Particle Physics, ETH Zurich, 8092 Zurich, Switzerland}

\ead{claudia.strabel@cern.ch}

\begin{abstract}
In this article the NA61/SHINE detector will be presented which is a large acceptance hadron spectrometer at the CERN SPS. 
It allows for a precise study of particle production from interactions of a 30~GeV proton beam in a carbon target 
in order to predict the neutrino flux of the T2K experiment at J-PARC, Japan. Requirements for the T2K experiment will be 
discussed together with the ongoing NA61 measurements. In particular preliminary NA61 results on pion production and cross 
section measurements from the 2007 pilot run will be shown.
\end{abstract}
\parskip 12pt
The NA61/SHINE (SHINE = SPS Heavy Ion and Neutrino Experiment) experiment at the CERN SPS combines a rich physics program 
in various fields~\cite{intro1,intro2,intro3,run2007}. Besides performing measurements for the T2K experiment, it takes a variety of data used for the description 
of cosmic-ray air showers in the Pierre Auger and KASCADE experiments~\cite{Auger,KASCADE} as well as for studying the behaviour of strongly 
interacting matter at high density. This article focuses on the NA61 hadron production measurements in p+C interactions 
which are targeted for an accurate neutrino flux prediction of the T2K experiment~\cite{t2k}.

T2K is a long baseline neutrino oscillation experiment at J-PARC, Japan, with the aim to precisely measure the 
$\nu_{\mu}\rightarrow \nu_{e}$ appearance and $\nu_{\mu}$ disappearance.
To generate neutrinos a high intensity 30~GeV proton beam impinging on a 90~cm 
long carbon target is used, whereby mesons ($\pi$, $K$) are produced which decay into neutrinos ($\nu_{\mu ,e}$). 
The neutrino flux is then measured 2.5$^\circ$ off-axis both in a near detector,
located 280~m behind the target, and the 295~km far away Super-Kamiokande detector (SK).
Neutrino oscillations can be probed by comparing the neutrino flux measured at SK to the one predicted at SK.
In order to predict the flux at SK  
one uses the near detector measurements  
and extrapolates them  
with the help of Monte Carlo (MC) predictions to SK. Up to now, these MC predictions depend on hadron production models.
For more precise predictions measurements of pion and kaon production off the carbon target are essential. 
The aim of the NA61/SHINE measurements for T2K is to provide this information.

\begin{figure}[!h]
\begin{center}
\includegraphics[scale=0.5]{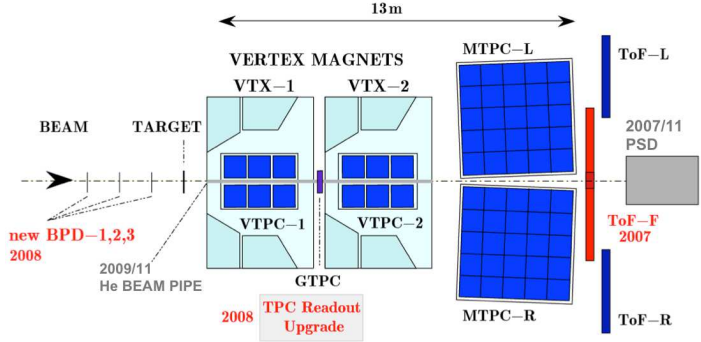}
\end{center}
\caption{The layout of the NA61/SHINE set-up (top view, not in scale).}
\label{Detector}      
\end{figure}

The NA61 detector is a large acceptance hadron spectrometer at the CERN SPS~(see Fig.~\ref{Detector}). 
It's main components
originate from the NA49 experiment~\cite{na49-nim}. These include four large volume time projection chambers
(TPCs), which 
provide high momentum resolution ($\sigma(p)/p^2 \cong (0.3-7)\cdot10^{-4}$ (GeV/c)$^{-1}$) and 
good particle identification (PID) via specific energy loss measurements. The PID capability of the TPCs is
complemented through velocity measurements of two time-of-flight (ToF-L/-R) detector arrays with a time measurement
resolution of $\sigma \approx 60$~ps. In 2007 an additional ToF detector (ToF-F) with a resolution of 
$\sigma \approx 120$~ps was installed in order to extend the PID acceptance to low momenta 
($p < 4$~GeV/c) as required for T2K~\cite{run2007}. 
Two carbon (isotropic graphite, $\rho = 1.84$ g/cm$^{3}$) targets of different lengths are used:
\begin{itemize}
\item a 2~cm long target ($\sim$4\% of nuclear interaction length, $\lambda_{I}$), the so called \textbf{thin target}, and
\item a 90~cm long cylinder of 2.6~cm diameter ($\sim$1.9 $\lambda_{I}$), the so called \textbf{T2K replica target}.
\end{itemize}
Proton beam particles are identified and selected by means of CERDAR and threshold Cherenkov
counters as well as several scintillation counters.
Interactions in the thin target are selected by an anti-coindidence of the incoming beam particle 
with a small scintillation counter (S4) of 2~cm diameter placed on the beam axis between the two vertex magnets.

\begin{figure}[!ht]
\begin{center}
\includegraphics[scale=0.5]{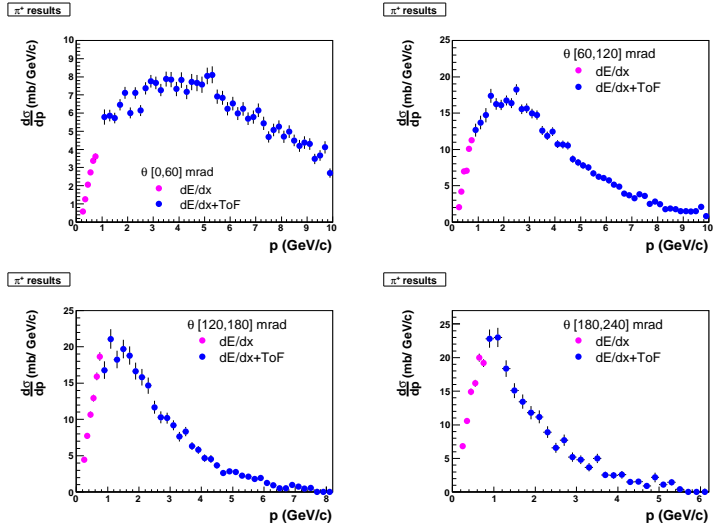}
\end{center}
\caption{Differential inclusive inelastic cross sections for $\pi^{+}$ versus momentum in different $\theta$ angle intervals
from dE/dx (pink) and dE/dx + ToF (blue) analyses.}
\label{piplus}
\end{figure}

\begin{figure}[!ht]
\begin{center}
\includegraphics[scale=0.5]{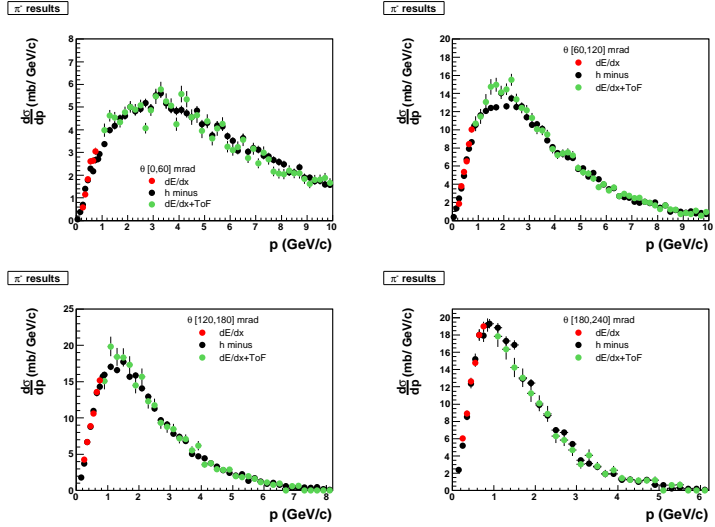}
\end{center}
\caption{Differential inclusive inelastic cross sections for $\pi^{-}$ versus momentum in different $\theta$ angle intervals
from dE/dx (red) and dE/dx + ToF (green) and $h^{-}$ (black) analyses.}
\label{pimin}
\end{figure}

During the pilot run~\cite{run2007} in October 2007 approximately 660k events with the thin target (target in), 
80k events without target (target out)
and 220k events with the T2K replica target 
were registered with a 30~GeV proton beam. From the observed interaction probabilities for target in ($P_{T_{in}}$) and 
out ($P_{T_{out}}$) configurations a trigger cross section ($\sigma_{trig}$) has been evaluated as:
\begin{equation}
    \label{eq:trig}
\sigma_{trig}  = \frac {1} {\rho L_{eff} N_{A}/A} ~ (P_{T_{in}}-P_{T_{out}})=(297.5\pm0.7 (stat.)\pm<10\%(syst.)\footnote{A more precise evaluation of the systematic
error is under investigation.} )~{\mathrm{mb}}~~~,
\end{equation}
where $N_A$, $\rho$, $A$ and $L_{eff}$ denote, respectively,
the Avogadro constant, the target density, it's atomic number and it's effective length.
The inelastic cross section ($\sigma_{inel}$) can be obtained from $\sigma_{trig}$
by applying two major corrections accounting for trigger biases. The first one is to subtract
the elastic contribution due to large angle coherent scattering of primary protons which 
do not reach the trigger counter S4 even though no inelastic interaction occured.
The second one is to add the lost inelastic contribution coming from secondary particles hitting
S4 and therefore preventing from triggering on the event.
Both corrections have been
estimated with GEANT4 simulations and result in $\sigma_{inel} = (254.7\pm1.0(stat.)\pm<10\%(syst.)\footnotemark[\value{footnote}])$~mb.
$\sigma_{trig}$ and $\sigma_{inel}$ are finally used for the normalization of the particle spectra. Up to now, preliminary
spectra for positively and negatively charged pions have been obtained via three different particle identification methods based on:

\begin{figure}
\begin{center}
\mbox{
\includegraphics[angle=270,scale=0.2]{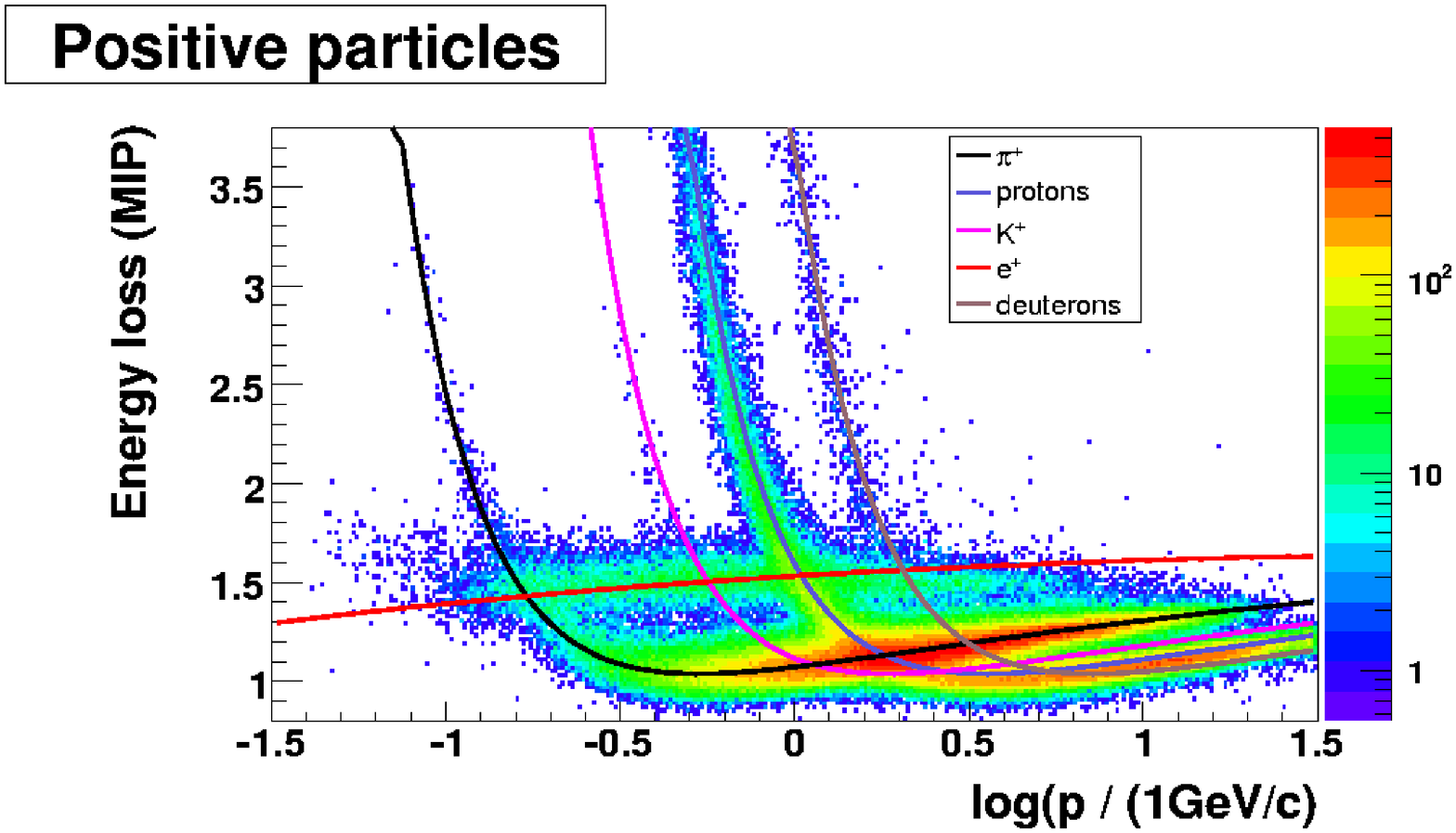}
\includegraphics[angle=270,scale=0.2]{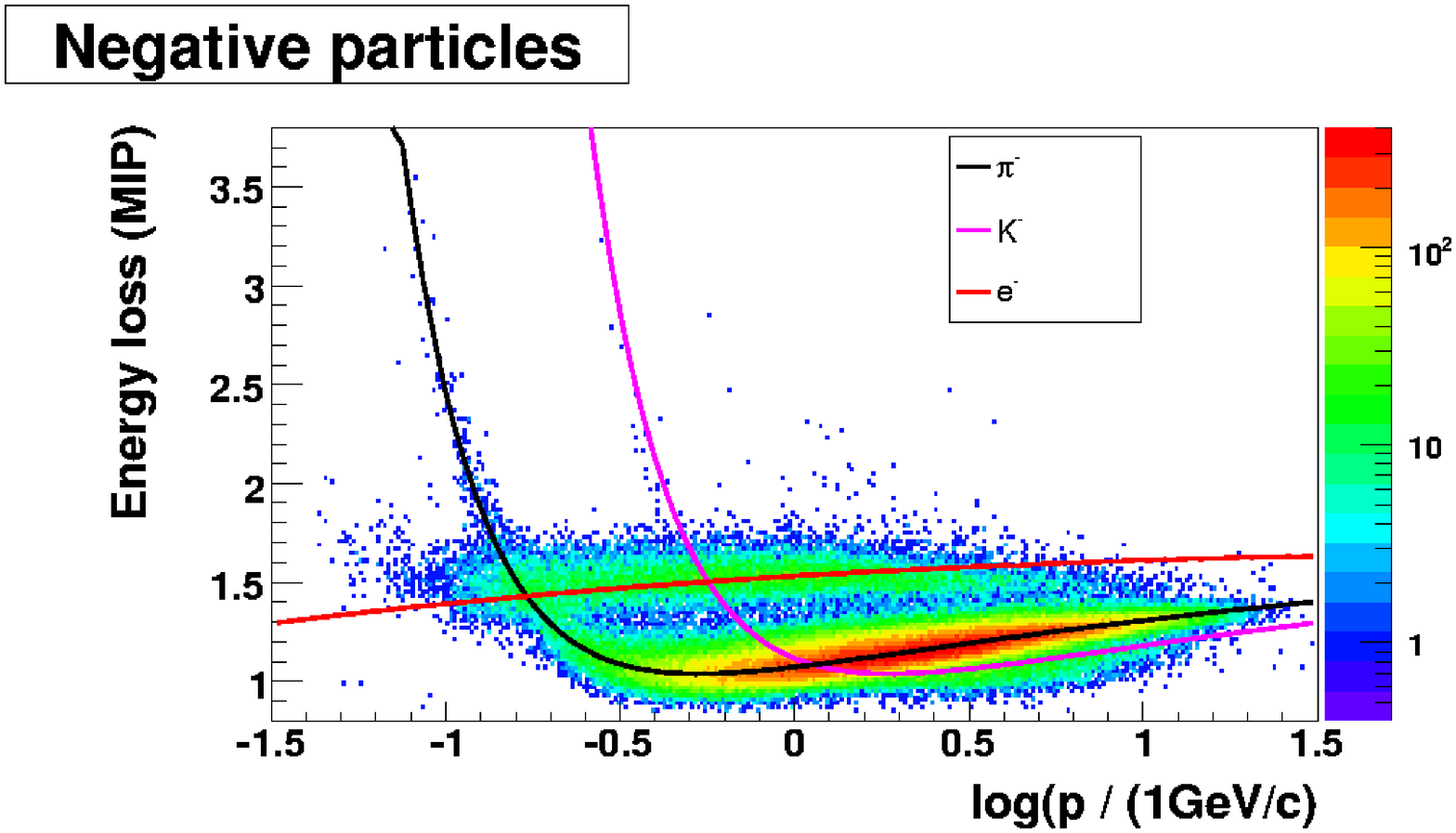}  
}
\mbox{
\includegraphics[scale=0.2]{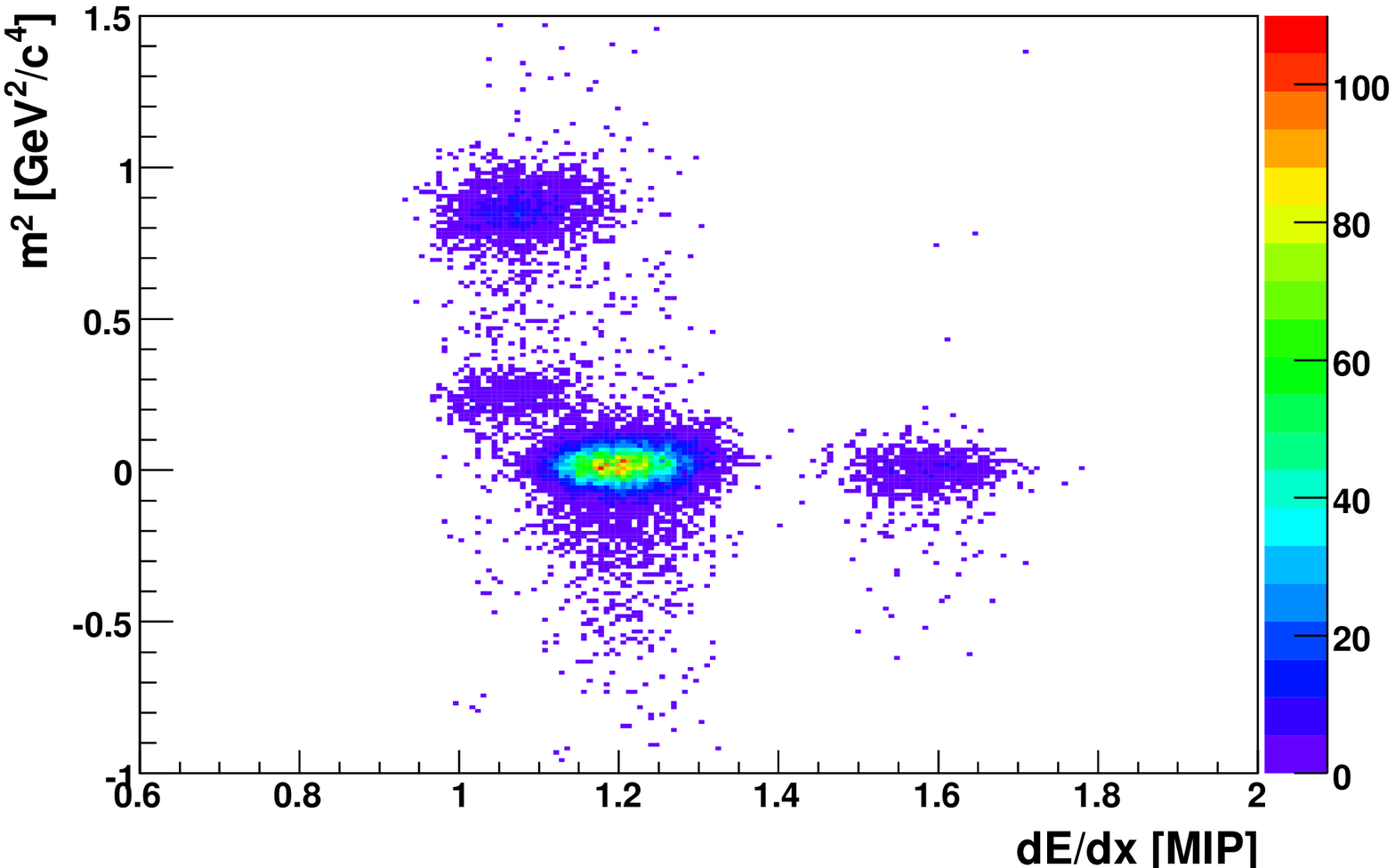}
\includegraphics[scale=0.2]{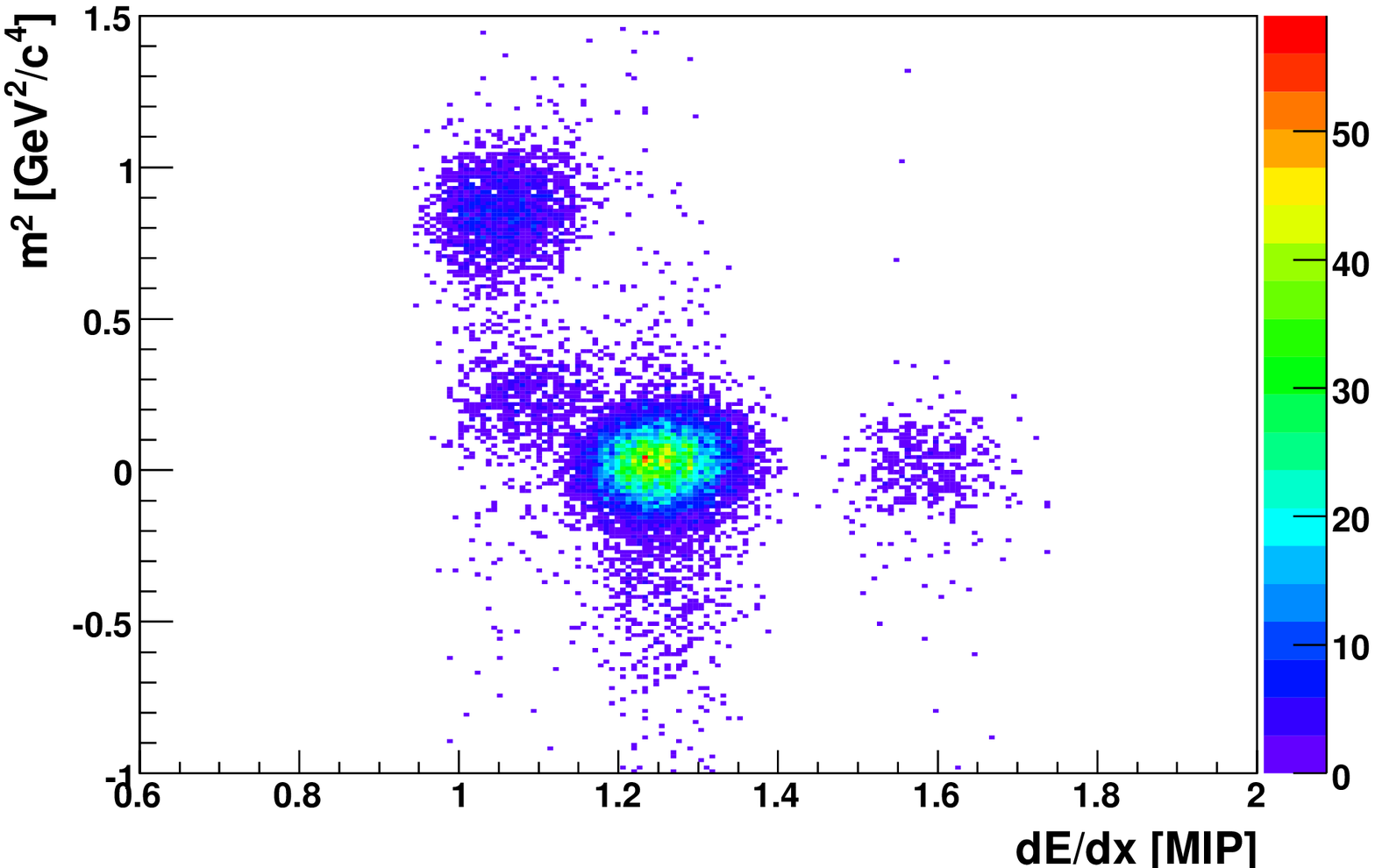} 
\includegraphics[scale=0.2]{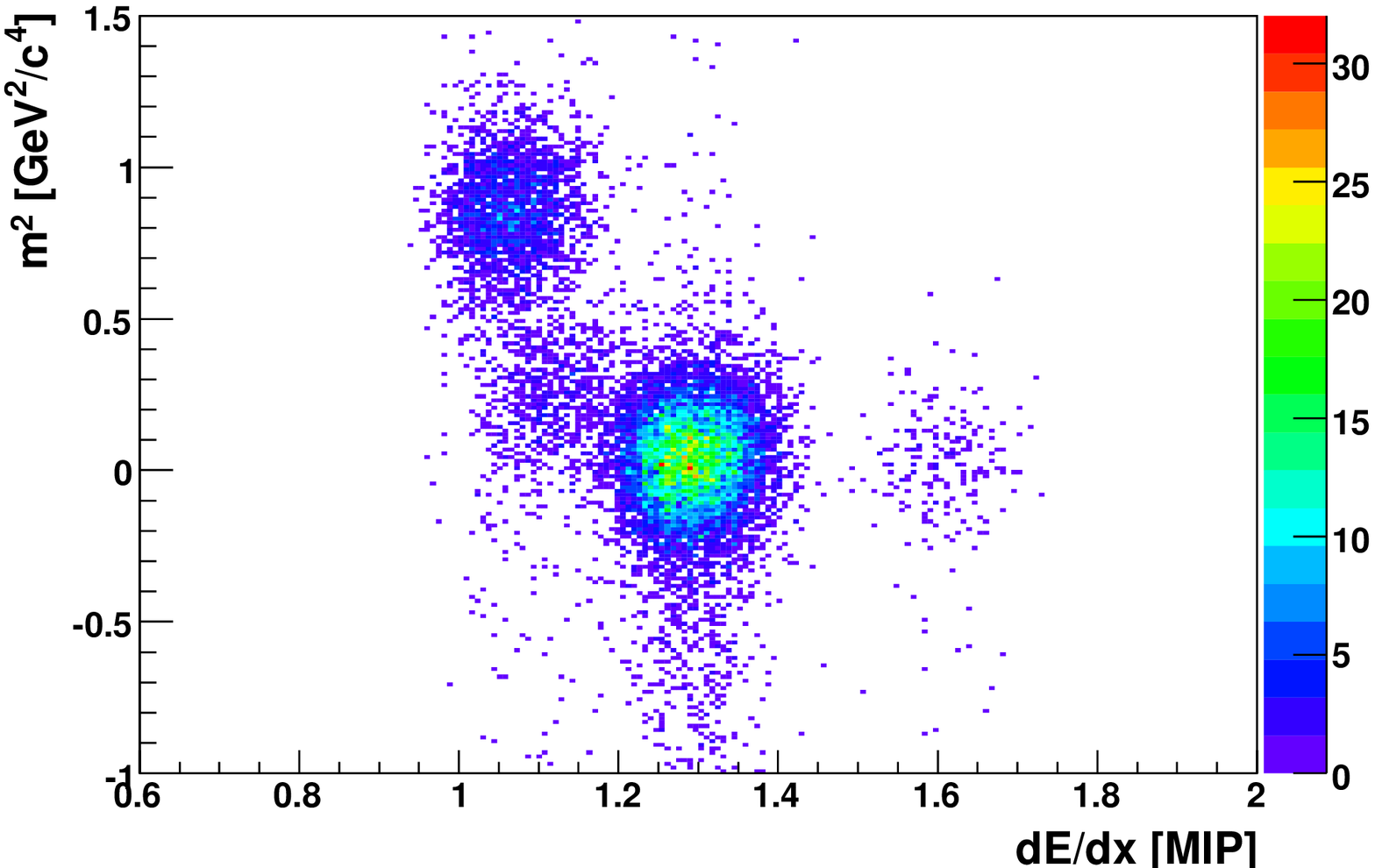} 
}
\end{center}
\caption{{\it Upper panel}: dE/dx as a function of momentum for p+C at 31~GeV/c beam momentum for positively and negatively 
charged particles, respectively, together with the Bethe-Bloch curves.
{\it Lower panel}: Examples of combined ToF and dE/dx PID performance for positively charged particles in the momentum range
2-3 GeV/c (left), 3-4 GeV/c (middle) and 4-5 GeV/c (right).} 
\label{dedx}
\end{figure}
\begin{itemize}
\item \textbf{Energy loss measurements}
in the TPC gas~(see Fig.~\ref{dedx}). A dedicated dE/dx analysis
for particles with momentum below 1~GeV/c, which is the region not covered by the time-of-flight detectors, was performed.  
\item \textbf{Combined energy loss and time-of-flight measurement} information (dE/dx + ToF). It was used to 
perform identification of pions with momenta above 1 GeV/c (see Fig.~\ref{dedx}).
\item \textbf{Analysis of negatively charged particles}
further referred to as $h^{-}$. It is based on the theoretical and experimental premises that negative particles produced 
by 30~GeV protons consist mainly of negative pion mesons with a small admixture of electrons, negative kaons and a negligible 
fraction of antiprotons. This procedure allows to obtain spectra of $\pi^{-}$ mesons in the full momentum range.
\end{itemize}
For the three different PID methods a common Venus-GHEISHA and Geant Monte Carlo simulation chain is used 
to calculate corrections for geometric acceptance, 
reconstruction efficiency, weak decays and vertex association.
Preliminary results on $\pi^{+}$ and $\pi^{-}$ differential inclusive inelastic cross sections 
obtained from the different
analyses are presented in Figs.~\ref{piplus} and~\ref{pimin}, respectively. Only statistical
errors are shown. The spectra obtained using the different methods agree better than by
20\%, which yields the upper limit for systematic biases. Work to minimize these biases is 
currently in progress.
Measurements obtained with the T2K replica target are currently being analysed. In 2009
further 6 million interaction triggers for the thin target and 4 million for the T2K replica
target were recorded, which will increase the limited statistics from the 2007 run by a 
factor of about 10.

\end{document}